\documentstyle[12pt]{article}
\makeatletter
\def\eqalign#1{\null\vcenter{\def\\{\cr}\openup\jot\m@th
  \ialign{\strut$\displaystyle{##}$\hfil&$\displaystyle{{}##}$\hfil
      \crcr#1\crcr}}\,}
\makeatother
\def\la{\lambda}
\begin{document}
\bigskip
\bigskip
\bigskip
\begin{center}
{\Large\bf 
%Explicit solution to some second-order differential and $q$-difference 
%eigenvalue equations related to $sl_2$ and $U_q(sl_2)$
Quasi-exactly solvable problems and the dual (q-)Hahn polynomials
}\\
\bigskip
\bigskip
\bigskip
\bigskip
{\large I. V. Krasovsky}\\
\bigskip
Max-Planck-Institut f\"ur Physik komplexer Systeme\\
N\"othnitzer Str. 38, D-01187, Dresden, Germany\\
E-mail: ivk@mpipks-dresden.mpg.de\\
\medskip
and\\
\medskip
B.I.Verkin Institute for Low Temperature Physics and Engineering\\
47 Lenina Ave., Kharkov 310164, Ukraine.
\end{center}
\bigskip
\bigskip
\bigskip

\noindent
{\bf Abstract.}
A second-order differential (q-difference) eigenvalue equation is constructed
whose solutions are generating functions of the dual (q-)Hahn
polynomials. The fact is noticed that these generating functions are reduced to
the (little q-)Jacobi polynomials, and implications of this for quasi-exactly
solvable problems are studied.
A connection with the Azbel-Hofstadter problem is indicated.

\newpage
\section{Introduction.}
In the present paper we shall consider some 
second-order differential and $q$-difference eigenvalue equations 
of the form:  
\begin{equation}
a(z)f^{''}(z)+b(z)f^{'}(z)+c(z)f(z)=\la f(z)
\end{equation}
and 
\begin{equation}
\alpha(z)f(q^sz)+\beta(z)f(q^{s+1}z)+\gamma(z)f(q^{s+2}z)=\la f(z)
\end{equation}
respectively, where the functions $a(z)$, $b(z)$, $c(z)$ are polynomials 
in $z$, while $\alpha(z)$, $\beta(z)$, $\gamma(z)$, in $z$ and $z^{-1}$. 
We shall be looking for polynomial
solutions $f(z)$. Note that using a transformation of the type
$\psi(y)=g(y)f(z(y))$ we can always reduce (1) to the Schr\"odinger form
$-\frac{d^2}{dy^2}\psi(y)+V(y)\psi(y)=\la\psi(y)$.

After subjecting the coefficients of (1) and (2) to certain general 
conditions, we shall see that the polynomial solutions to equations 
(1) and (2) are given by generating functions of the dual Hahn and dual 
$q$-Hahn polynomials, respectively. These solutions are explicit in the sense 
that all the eigenvalues $\la$
and corresponding eigenfunctions $f(z)$ in the space of polynomials of degree
at most $N$ are known explicitly.
 
Let us first consider equation (1). It is known
[\ref{Turbiner0}] (see also [\ref{Turbiner}] for a recent review of the 
subject) that the spectral
problem for the operator $D=a(z)\frac{d^2}{dz^2}+b(z)\frac{d}{dz}+c(z)$
in the space ${\cal H}_N$ spanned by the vectors ${1,z,z^2,\dots,z^N}$
is closely related to the representation theory of the algebra $sl_2$.
${\cal H}_N$ is a representation space of this algebra, and the generators 
of $sl_2$ have in this representation the following form:
\begin{equation}
J^+=z^2\frac{d}{dz}-Nz;\qquad J^0=z\frac{d}{dz}-N/2;
\qquad J^-=\frac{d}{dz}.
\end{equation}
The necessary and sufficient condition [\ref{Turbiner2}]
for the operator $D$ to leave ${\cal H}_N$ invariant is that $D$ be 
expressed in the form
\begin{equation}
\eqalign{
D=c_{++}J^+J^+ + c_{+0}J^+J^0 + c_{+-}J^+J^- +  c_{0-}J^0J^- +\\
c_{--}J^-J^- + c_{+}J^+ + c_{0}J^0 + c_{-}J^- + d,}
\end{equation}
where $c_{i,j}, c_{i}, d$ are constant parameters.
Henceforth, we shall assume that (4) holds. According to the  
classification of Turbiner, 
equation (1) is called in this case quasi-exactly-solvable.\footnote{
For the expressions of all possible potentials in the corresponding 
Schr\"odinger equations see [\ref{Gonzalez}].}

If the parameters $c_{++}=c_{+0}=c_{+}=0$ then $D$ is, obviously, upper
diagonal in the basis of monomials $\{z^k\}_{k=0}^N$, and hence, it preserves 
the flag ${\cal H}_0\subset{\cal H}_1\subset\cdots\subset{\cal H}_N$.
The coefficient $a(z)$ in this case is a polynomial of no
more than the second degree, $b(z)$, the first degree, and $c$ is independent
of $z$. Hence, the operator $D$ preserves ${\cal H}_N$ for any $N$.
Corresponding equation (1) is then called exactly-solvable\footnote{
An important algebraic approach to the exactly solvable problems is
proposed in [\ref{Granovskii}].}.
It is easy to verify that in this case, changing the six remaining parameters
$c_{+-}$, $c_{0-}$, $c_{--}$, $c_{0}$, $c_{-}$, $d$, we can obtain for any $N$
an arbitrary operator $D$ with the just mentioned restriction on the 
degrees of $a(z)$, $b(z)$, and $c(z)$. 
The full classification of the polynomial solutions to (1) for such an
operator $D$ is available in the literature (e.g., [\ref{Chihara}]).
All these solutions can be written in an explicit form.
In particular, the classical orthogonal polynomials (Jacobi, Laguerre, 
and Hermite) satisfy exactly solvable equations of the type (1).  
In the present paper, however, 
we shall be interested in a different type of solutions.

Take once again the general case (4) and let now $c_{++}=c_{--}=0$.
Then it is seen from (3) and (4) that the operator $D$ is represented in the 
basis of monomials $\{z^k\}_{k=0}^N$ by a tridiagonal matrix. The equation (1)
takes on the following form ($f(z)=\sum_{k=0}^Nz^kp_k$)\footnote{
The quantity $\sum_{k=0}^N\alpha_kz^kp_k$, where $\alpha_k$ are some
parameters, is called a generating function of the sequence 
$\{p_k\}_{k=0}^N$.}:
\begin{equation}
\pmatrix {
a_0-\la & b_0 &&&& 0\cr
c_1 & a_1-\la & b_1\cr
& c_2 & a_2-\la & b_2\cr
&& \ddots & \ddots & \ddots\cr
0 &&&& c_N & a_N-\la }
\pmatrix {p_0\cr p_1\cr\vdots\cr\vdots\cr p_N}
=0
\end{equation}
We see that the quantities $p_k$ satisfy the three-term recurrence relation
\begin{equation}
c_k p_{k-1}+(a_k-\la)p_k+b_k p_{k+1}=0,\qquad  p_{-1}=0.
\end{equation}
Thus [e.g., \ref{Chihara}], $p_k(\la)$ form
a finite system of orthogonal polynomials.
(See [\ref{UZ},\ref{Zaslavskii},\ref{BD},\ref{Finkel}] for studies related 
to this aspect of quasi-exact solvability.)

The matrix elements $a_k$, $b_k$, $c_k$ are polynomials of degree 2 in the
index $k$. The coefficients of these polynomials are expressed in terms of the
7 free parameters $c_{+0}, c_{+-}, c_{0-}, c_+, c_0, c_-, d$.
It is easy to verify that $a_k$, $b_k$, $c_k$ can be obtained in the following
way. Assume  $A(k)$, $B(k)$, $C(k)$ to be arbitrary polynomials in $k$ of
degree 2 and impose the boundary conditions $C(N+1)=B(-1)=0$.
Then $a_k=A(k)$, $b_k=B(k)$, $c_k=C(k)$, $k=0,\dots,N$.

Now consider a particular case of (5). Namely, impose the Askey-Wilson
condition for the transposed matrix $a_k+b_{k-1}+c_{k+1}=0$ and the
restriction
$c_{+0}=c_{0-}$. Then the polynomials 
$\{\hat p_k(\lambda/c_{+0})\}_{k=0}^N$ associated with the transposed matrix,
are the dual Hahn hypergeometric orthogonal polynomials.
One of their generating functions provides an explicit solution to (1).
In Section 2 we show that the corresponding equation (1) is reduced to 
the exactly
solvable (see above) equation for the Jacobi polynomials by means of the 
transformation $\psi(y)=(y+1)^Nf(\frac{y-1}{y+1})$. (Note that this is a
particular case of the transformation that connects various forms of a
quasi-exactly solvable equation [\ref{Finkel}]. Such transformations 
comprise the irreducible representation of the group $GL(2)$ in the space of
polynomials.)   
Equation (1) in this case has an infinite number of formal explicit 
solutions $f(z)$, but only $N+1$ of them are guaranteed to be polynomials.  

The above considerations can be generalized for equation (2).
Equation (2) is related to quantum deformations of the $sl_2$ algebra in a 
similar way as equation (1)
is related to $sl_2$ [\ref{Turbiner},\ref{WZ}]
\footnote{In addition to the results for differential and $q$-difference
equations reviewed in [\ref{Turbiner}], it is also possible [\ref{ST}] 
to obtain similar results for difference equations of the form
\mbox{$\sum_iA_i(x)f(x+\delta_i)=\lambda f(x)$}.}.
For connections between (quantum) groups and orthogonal polynomials 
see [\ref{VK},\ref{Koo}]. 

Let $\alpha(z)$, $\beta(z)$, $\gamma(z)$ in (2) be the first order polynomials
in $z$ and $z^{-1}$. Then, obviously, in the basis of monomials
$\{z^k\}_{k=0}^N$, equation (2) has the form (5) where  $a_k$, $b_k$, $c_k$
are expressed in terms of the coefficients of the polynomials 
$\alpha(z)$, $\beta(z)$, $\gamma(z)$. For the space spanned by
$\{z^k\}_{k=0}^N$ to be an eigenspace, these coefficients 
should be such that
$b_{-1}=c_{N+1}=0$. This condition implies that only 7 out of total 9 
coefficients are independent. We shall fix 3 more by requiring the
Askey-Wilson condition for the transposed matrix to hold:
$a_k+b_{k-1}+c_{k+1}=0$. Fixing then one of the remaining 4 free parameters 
in an appropriate way and putting $s=0$, we obtain the dual q-Hahn 
basic hypergeometric polynomials  
$\{\hat p_k(\lambda/\epsilon)\}_{k=0}^N$ as a system generated by the
recurrence relation 
$b_{k-1}\hat p_{k-1}+(a_k-\la/\epsilon)\hat p_k+c_{k+1}\hat p_{k+1}=0$.

In Section 3 we shall
consider equations of the type (2) whose solutions will be given 
by the generating functions of the dual $q$-Hahn
polynomials. These equations are related to the
$q$-difference equation for the little $q$-Jacobi polynomials.

It is interesting to note that the zeros of polynomial solutions of 
equations (1) and (2) are connected with the eigenvalues $\la$ by
a set of Bethe-ansatz type algebraic equations [\ref{Shifman},\ref{WZ}].

Thus, the message of the present paper can be summarized as follows.
The dual Hahn (dual $q$-Hahn) polynomials are the most general system in 
the Askey-scheme of the known hypergeometric 
(basic hypergeometric) orthogonal polynomials [\ref{Koekoek}] whose 
generating function provides an explicit polynomial solution to the 
eigenvalue equation (1) (equation (2)). This generating function of
the dual Hahn (dual $q$-Hahn) polynomials is reduced to the Jacobi (little 
$q$-Jacobi) polynomials. 
(The corresponding problems are, thus, exactly solvable.)

\section{Dual Hahn polynomials and a differential equation}
 
Unlike in the introduction, we shall now begin with the dual Hahn polynomials
rather than with the differential equation.

The dual Hahn polynomials are defined by the formula (e.g.,[\ref{Koekoek}])
\begin{equation}
p_n(\la(x))=\sum_{k=0}^n\frac{(-n)_k(-x)_k(x+\gamma+\delta+1)_k}
{(\gamma+1)_k(-N)_k k!},\qquad  n=0,1,\dots,N,\label{def}
\end{equation}
where $\gamma$ and $\delta$ are fixed parameters and the ``shifted'' 
factorial
is defined as $(a)_0=1$, $(a)_k=a(a+1)\cdots(a+k-1)$, $k=1,2,\dots$. 
The polynomials (\ref{def}) satisfy the recurrence relation (which we will 
formally consider for an arbitrary integer $n$)
\begin{equation}
\eqalign{
\la(x)p_n=A_np_{n+1}-(A_n+C_n)p_n+C_np_{n-1},\\
A_n=(n-N)(n+\gamma+1),\qquad C_n=n(n-\delta-N-1),\\
\la(x)=x(x+\gamma+\delta+1).}\label{r1}
\end{equation}

The above three-term recurrence relation can be viewed as the
eigenvalue equation for an infinite tridiagonal matrix $M$, $p_n$'s
being components of an eigenvector. For what follows, we would need to
demand that the finite dimensional space ${\cal L}_N$ corresponding to
the indices $n=0,1,\dots,N$ be invariant under the action of the matrix
$M$. This would be the case if the matrix elements
$M_{-1\,0}=M_{N+1\,N}=0$. Since for our matrix $M_{0\,-1}=C_0=0$ and
$M_{N\,N+1}=A_N=0$, the transposed matrix $M^{\small T}$ will have the
desired property of preserving ${\cal L}_N$. The polynomials associated with
$M^{\small T}$ satisfy the recurrence

\begin{equation}
\la(x){\tilde p}_n=C_{n+1}{\tilde p}_{n+1}-
(A_n+C_n){\tilde p}_n+A_{n-1}{\tilde p}_{n-1};\label{r2}
\end{equation}
and it is easy to show by induction that
\[
{\tilde p}_n=\frac{A_0A_1\cdots A_{n-1}}{C_1C_2\cdots C_n}p_n=
\frac{(-N)_n(\gamma+1)_n}{(-\delta-N)_n n!}p_n.
\]
Now multiply both sides of (\ref{r2}) by $z^n$ and perform summation 
over $n$ from $n=0$ to $N$. We obtain
\begin{equation}
\eqalign{
\la f(z)=z(z-1)^2 f^{''}(z)+\\
\{(\gamma-N+2)z^2-(\gamma-\delta-2N+2)z-
(\delta+N)\}f^{'}(z)-\\
N(\gamma+1)(z-1)f(z),}\label{de}
\end{equation}
where $f(z)=\sum_{n=0}^N z^n{\tilde p}_n$. To get the {\it homogenious} 
equation (\ref{de}), it was necessary to put
${\tilde p}_{-1}=0$ and ${\tilde p}_{N+1}=0$ (we can do this because we are 
looking for solutions in ${\cal L}_N$). 

We can represent (\ref{de}) in the form $\la f(z)=Df(z)$ as the eigenvalue 
equation for a second-order differential 
operator $D$ in the space ${\cal H}_N$ spanned by monomials $\{z^k\}_{k=0}^N$.
One can already notice that (\ref{de}) can be reduced to a hypergeometric
equation. However, we shall follow another approach which can be easier
generalized to $q$-difference equations. 

Since $M^{\small T}$ in  ${\cal L}_N$ is just the matrix representation 
of the operator $D$ in the basis  $\{z^k\}_{k=0}^N$, 
the eigenvalues of $D$ in ${\cal H}_N$ and 
$M^{\small T}$ in ${\cal L}_N$ are the same. 
To find them, first replace the parameter $N$ in (\ref{def}) and (\ref{r1}) by
$N+\epsilon$, $\epsilon\neq 0$. Then (\ref{def}) will be valid not only for
$n=0,1,\dots,N$, but also for $n=N+1$. We find the eigenvalues from the
equation:
\begin{equation}
0=\det(M^{\small T}-\la I)=\det(M-\la I)=
\lim_{\epsilon\to 0}A_0 A_1\cdots A_N p_{N+1}(\lambda).\label{sec}
\end{equation}
Here (only) one of the factors $A_i$ goes to zero as $\epsilon\to 0$:
$A_N=-\epsilon(N+\gamma+1)$. Furthermore, only the addend with the index 
$k=N+1$ in the expression 
\[
p_{N+1}(\la(x))=\sum_{k=0}^{N+1}\frac{(-N-1)_k(-x)_k(x+\gamma+\delta+1)_k}
{(\gamma+1)_k(-N-\epsilon)_k k!}
\]
is not bounded as $\epsilon\to 0$ (growing as $1/\epsilon$). 
Hence (\ref{sec}) is equivalent to
$(-x)_{N+1}(x+\gamma+\delta+1)_{N+1}=0$.
From here, using the definition of $\la$ in (\ref{r1}), we obtain 
the eigenvalues:
\begin{equation}
\la(m)=m(m+\gamma+\delta+1),\qquad m=0,1,\dots,N.\label{eigval1}
\end{equation}
The corresponding eigenvectors of $D$ are
\[
f_m(z)=\sum_{n=0}^N z^n \frac{(-N)_n(\gamma+1)_n}{(-\delta-N)_n n!}
p_n(\la(m))
\]
Notice that this generating function is one of those admitting representation
in terms of the hypergeometric series [\ref{Koekoek}]:
\begin{equation}
f_m(z)=
(1-z)^m\sum_{k=0}^{N-m}\frac{(m-N)_k(m+\gamma+1)_k}{(-\delta-N)_k k!}z^k.
\label{eigvect}
\end{equation}
Using one of the representations of the Jacobi polynomials 
(see, e.g., [\ref{AS}]) we can rewrite (\ref{eigvect})
in the form:
\begin{equation}
f_m(z)=\frac{(N-m)!}{(-N-\delta)_{N-m}}(1-z)^N 
P^{(-\delta-N-1,-\gamma-N-1)}_{N-m}\left(\frac{1+z}{1-z}\right),\label{J}
\end{equation}
where $P^{(\alpha,\beta)}_k(x)$ is the usual notation for the Jacobi
polynomial. Thus, this formula expresses the generating function of the dual
Hahn polynomials in terms of the Jacobi polynomials.
It is easy to transform (\ref{de}) into the
hypergeometric equation for the Jacobi polynomials.
Since the equation for the Jacobi polynomials is valid for an
arbitrary large index $k$, formula (\ref{J}) provides an infinite number of
{\it nonpolynomial} solutions to (\ref{de}) for $m=-1,-2,\dots$. 

The operator $D$ is expressed in the following form in terms of the 
generators (4): 
\begin{equation}
\eqalign{
D=J^+J^0 -2J^+J^- +J^0J^- +(\gamma+1+N/2)J^+ +(\delta-\gamma-2)J^0 -\\
(N/2+\delta)J^- +N(\delta+\gamma)/2}
\end{equation}

After the transformation
\begin{equation}
\psi(y)=f(\coth^2 {y\over 2})
\left\{\sqrt{\sinh y}\sinh^\gamma{y\over 2}\cosh^\delta{y\over 2}
\coth^N{y\over 2}\right\}^{-1}
\end{equation}
equation (\ref{de}) is reduced to the Schr\"odinger-type equation
\begin{equation}
\eqalign{
-\frac{d^2}{dy^2}\psi(y)+V(y)\psi(y)=\varepsilon\psi(y),\\
V(y)=\frac1{2\sinh^2y}\left\{(\gamma-\delta)(2N+\gamma+\delta+2)\cosh y+ 
\right.\\
\left.(N+\gamma)^2+(N+\delta)^2+2(2N+\gamma+\delta)+{3\over2}\right\}+
{1\over4}(1+\gamma+\delta)^2,}\label{se}
\end{equation}
with formal solutions:
\begin{equation}
\eqalign{
\varepsilon_m=-m(m+\gamma+\delta+1),\qquad m=\dots,-1,0,1,\dots,N,\\
\psi_m(y)=\frac{c_m P^{(-\delta-N-1,-\gamma-N-1)}_{N-m}(-\cosh y)}
{\sqrt{\sinh y}\sinh^\gamma{y\over 2}\cosh^\delta{y\over 2}
\coth^N{y\over 2}(1-\cosh y)^N},}\label{e}
\end{equation}
where $c_m$ is a constant factor.
This is the exactly solvable Schr\"odinger equation related to the Jacobi
polynomials.
It is easy to verify that if $\gamma+N<0$ and $2m+\gamma+\delta+1>0$, then
the function $\psi_m(y)$ belongs to the space $L^2(-\infty,\infty)$ (that is
$\int_{-\infty}^{\infty}|\psi_m|^2dy<\infty)$. In this case, since the 
operator $-d^2/dy^2 +V(y)$ is symmetric with respect to the inner product
$(f,g)=\int_{-\infty}^{\infty}f(y)\overline{g(y)}dy$, such eigenfunctions 
$\psi_m(y)$ corresponding to different $\varepsilon_m$ are orthogonal.

Note that, generally, $\psi_m(y)$ and $\psi_m'(y)$ are discontinuous at $y=0$.
Consider the physically more reasonable Schr\"odinger equation (\ref{se}) in
the space $L^2(0,\infty)$ with the boundary condition $\psi(0)=0$.
The functions $\psi_m(y)$ for $m=0,1,\dots, N$ belong to this space if 
$\gamma+N<-1/2$, $\gamma+\delta+1>0$. 
The corresponding $\varepsilon_m$ are the levels of the discrete spectrum
because they are less then the asymptotics 
$(1+\gamma+\delta)^2/4$ of the potential as $y\to\infty$; and
$V(y)$ goes to $-\infty$ as $y\to 0$.
Note that $\varepsilon_m$, $m=N,\dots,1,0$ are the first $N+1$ lowest 
eigenvalues of the Schr\"odinger operator.

\section{Continuous dual $q$-Hahn (dual $q$-Hahn) 
polynomials and a $q$-difference equation}
\subsection{$q$ a root of unity}
The continuous dual $q$-Hahn polynomials (which depend on parameters $a$, $b$,
and $c$) are defined by the expression
(e.g.,[\ref{Koekoek}])
\begin{equation}
p_n(x)=\sum_{k=0}^n\frac{(q^{-n};q)_k(at;q)_k(at^{-1};q)_k q^k}
{(ab;q)_k(ac;q)_k(q;q)_k},\qquad 2x=t+t^{-1},
\quad n=0,1,\dots,\label{defq}
\end{equation}
where $(d;q)_0=1$ and $(d;q)_k=(1-d)(1-dq)\cdots(1-dq^{k-1})$, $k=1,2,\dots$. 
(In fact the $n$'th continuous dual $q$-Hahn polynomial differs from $p_n(x)$
by a constant.)
They satisfy the recurrence relation
\begin{equation}
\eqalign{
2x p_n=A_np_{n+1}+(a+a^{-1}-A_n-C_n)p_n+C_np_{n-1},\\
A_n=a^{-1}(1-abq^n)(1-acq^n),\qquad C_n=a(1-q^n)(1-bcq^{n-1}).}\label{r1q}
\end{equation}

As in the previous section, introduce the matrix $M_q$ associated with the
eigenvalue problem (\ref{r1q}) and the space ${\cal L}_k$ corresponding to
the indices $n=0,1,\dots,k$.
Let us first consider the case when $q$ is an $N$'th primitive root of 
unity\footnote{Note that the basic hypergeometric polynomials for $q$ a root
of unity have a number of interesting properties and applications 
[\ref{SZ}].}, 
that is $q=e^{2\pi iS/N}$, where $S$ and $N$ are positive integers 
which do not have a common divisor other than $1$. 
Let us set furthermore $ac=q$. Then, obviously, $M_q$ preserves 
${\cal L}_{N-1}$. (Moreover, the orthogonal complement of ${\cal L}_{N-1}$
to the whole space where $M_q$ acts is also invariant with respect to $M_q$.)

Multiplying both sides of the recurrence relation (\ref{r1q}) by $z^n$ 
and performing summation from $n=0$ to $N-1$, we obtain
\begin{equation}
\eqalign{
2x f(z)=\{(az)^{-1}+az\}f(z)+\\
\{-(a^{-1}+bq^{-1})z^{-1}+a+2b+qa^{-1}-(a+b)qz\}f(qz)+\\
b\{(qz)^{-1}-q-1+q^2 z\}f(q^2 z),}\label{deq}
\end{equation}
where $f(z)=\sum_{n=0}^{N-1} z^n p_n$.

Proceeding in a similar way as in Section 2, we obtain the following set of 
solutions to (\ref{deq}) in the space spanned by $\{z^k\}_{k=0}^{N-1}$:
\begin{equation}
2x_m=aq^m+a^{-1}q^{-m},\label{eigenvalq}
\end{equation}
\begin{equation}
\eqalign{
f_m(z)=\sum_{n=0}^{N-1}z^n 
\sum_{k=0}^n\frac{(q^{-n};q)_k(a^2q^m;q)_k(q^{-m};q)_k q^k}
{(ab;q)_k((q;q)_k)^2}=\\
(qz;q)_{N-1-m}\sum_{k=0}^m \frac{(q^{-m};q)_k(ba^{-1}q^{-m};q)_k
(a^2 q^m z)^k}{(ab;q)_k (q;q)_k},\qquad m=0,1,\dots,N-1,}\label{eigenvecq}
\end{equation}
where in the last formula we used the equivalence of the continuous 
dual $q$-Hahn polynomials at
$ac=q^{-N+1}$ and the dual $q$-Hahn polynomials (to be verified below) and the
expression for a generating function of the dual $q$-Hahn polynomials 
[\ref{Koekoek}].

The solution is especially simple for $m=0$:
$2x_0=a+a^{-1}$, $f_0(z)=1+z+z^2+\cdots+z^{N-1}$. In this case we also know
explicitly the zeros of $f_0(z)$: $z_i=q^i$, $i=1,2,\dots, N-1$.
Note that the zeros of all $N$ solutions
$f_m(z)$ can be found in the case when $b=0$. Then it is a simple exercise
to obtain, using (\ref{deq}), the set of zeros 
$z(m)=\{z_1,z_2,\dots,z_{N-1}\}$ of  $f_m(z)$:
\begin{equation}
\eqalign{
z(m)=\{q^{m+1},q^{m+2},\dots,q^{N-1},a^{-2}q^{-m+1},a^{-2}q^{-m+2},\dots,
a^{-2}\},\\
m=1,2,3,\dots,N-2\\
z(0)=\{q,q^2,\dots,q^{N-1}\},\qquad z(N-1)=\{a^{-2}q^2,a^{-2}q^3,\dots,
a^{-2}q^N\}}
\end{equation}
 
The difference operator $D_q$ (defined by the equation (\ref{deq}) written in
the form $2xf(z)=D_qf(z)$) can be expressed in terms of the generators of the
$U_{q^{1/2}}(sl_2)$ algebra represented in ${\cal H}_{N-1}$. In a certain 
representation in
this space the generators have the form (we use the notation from [\ref{WZ}]):
\begin{equation}
\eqalign{
A=q^{-\frac{N-1}4}T_+,\qquad D=q^{\frac{N-1}4}T_-,\\
B=z(q^{1/2}-q^{-1/2})^{-1}(q^{\frac{N-1}2}T_- - q^{-\frac{N-1}2}T_+),\\
C=-z^{-1}(q^{1/2}-q^{-1/2})^{-1}(T_- - T_+),}
\end{equation}
(recall that $q=e^{2\pi i S/N}$) where the operators $T_+$ and $T_-$ act on a
vector $g(z)\in{\cal H}_{N-1}$ as follows: $T_\pm g(z)=g(q^{\pm 1/2}z)$.

As is easy to verify,
\begin{equation}
\eqalign{
D_q=A^2\{-b(1+q^{-1})A^2+ (q^{1/2}-q^{-1/2})(bq^{-\frac{N-1}4-1}CA+
aq^{-\frac{N-1}4-1}BD-\\
bq^{\frac{N-1}4}BA -a^{-1}q^{\frac{N-1}4+1}CD)+
(a+2b+a^{-1}q)q^{\frac{N-1}2}\}.}
\end{equation}

\subsubsection{Azbel-Hofstadter problem}
It was recently shown [\ref{WZprl}] that part of
the spectrum of the Hamiltonian in the Azbel-Hofstadter problem (of an 
electron on a square lattice subject to a perpendicular
uniform magnetic field) can be obtained as ($N$) solutions 
$\la$ of the following
equation in ${\cal H}_{N-1}$:
\begin{equation}
i(z^{-1}+qz)f(qz) - i(z^{-1}+q^{-1}z)f(q^{-1}z)=\la f(z),\label{H0}
\end{equation}
where $q=e^{i\Phi/2}$. $\Phi=4\pi S/N$ is the flux of the magnetic field per
plaquette of the lattice. (Henceforth, we assume that $N$ is odd.) The
spectrum has particularly interesting properties when
$S$, $N\to\infty$ so that $S/N\to\alpha$, where $\alpha$ is an irrational
number (see, e.g., [\ref{Hof},\ref{LJ}]).
Representation of (\ref{H0}) in the basis of monomials gives:
\begin{equation}
i(q^{n+1}-q^{-(n+1)})\tilde p_{n+1}+
i(q^n-q^{-n})\tilde p_{n-1}=\la \tilde p_n,
\qquad n=0,1,\dots,N-1,\label{H1}
\end{equation}
where the polynomials $\tilde p_n(\la)$ are defined by the formula
$f(z)=\sum_{n=0}^{N-1} z^n\tilde p_n(\la)$.

On the other hand, setting in (\ref{r1q}) $a=iq^{1/2}$ (hence,
$c=-iq^{1/2}$), $b=0$, we reduce (\ref{r1q}) to
\begin{equation}
\eqalign{
(1-q^{n+1})\hat p_{n+1}+
(1-q^n)\hat p_{n-1}=2x \hat p_n,\\
\hat p_n=a^{-n}p_n,
\qquad n=0,1,\dots,N-1,}\label{H2}
\end{equation}

If we denote the $N\times N$ matrices corresponding to eigenvalue equations
(\ref{H1}) and (\ref{H2}) by $H$ and $M$, respectively, then the following
expression holds:
\begin{equation}
H=(M-M^*)/i.\label{H3}
\end{equation}
In other words, $H$ is the imaginary part of $2M$. (Note that $M$ and its
adjoint $M^*$ do not commute.) The spectrum of $M$ is given by
(\ref{eigenvalq}) with $a=iq^{1/2}$: $2x_k=2\sin\frac{2\pi k}N$,
$k=0,1,\dots,N-1$.

Expression (\ref{H3}) provides a connection between the results of Section 3
and the Azbel-Hofstadter problem.

\subsection{Arbitrary $q$}
Equations (\ref{defq}) and (\ref{r1q}) are valid for an arbitrary complex
$q$ (except for certain fixed values which one can treat on the basis of
continuity considerations). In this general case, in order to 
obtain a $q$-difference equation
with the largest number of free parameters, we shall use the approach of
Section 2. Namely, consider the polynomials associated with the transposed
matrix $M_q^{\small T}$. Put $ac=q^{1-N}$, then the space ${\cal L}_{N-1}$
will be invariant with respect to $M_q^{\small T}$. (Note that unlike for
$q^{N}=1$, in the general case the orthogonal complement of ${\cal L}_{N-1}$
to the whole infinite-dimensional space is not invariant with respect to 
$M_q^{\small T}$.) The polynomials associated with $M_q^{\small T}$ are
connected with the dual $q$-Hahn polynomials as follows (c.f. Section 2):
\[
\tilde p_n=
\frac{(ab;q)_n(q^{-N+1};q)_n}{a^{2n}(q;q)_n(ba^{-1}q^{-N+1};q)_n}p_n.
\]
Proceeding as in Section 2, we obtain the following equation for the
generating function $f(z)=\sum_{n=0}^{N-1} z^n\tilde p_n$:
\begin{equation}
\eqalign{
2x f(z)=\{az^{-1}+a^{-1}z\}f(z)+\\
\{-(a+bq^{-N})z^{-1}+a+b+bq^{-N}+a^{-1}q^{-N+1}-(a^{-1}q^{-N+1}+b)z\}f(qz)+\\
bq^{-N}\{z^{-1}-q-1+q z\}f(q^2 z),}\label{deq2}
\end{equation}
Its solutions in the space spanned by $\{z^k\}_{k=0}^{N-1}$ are 
\begin{equation}
2x_m=aq^m+a^{-1}q^{-m},\qquad m=0,1,\dots,N-1,\label{eigenvalq2}
\end{equation}
\begin{equation}
\eqalign{
f_m(z)=\sum_{n=0}^{N-1}z^n 
\frac{(ab;q)_n(q^{-N+1};q)_n}{a^{2n}(q;q)_n(ba^{-1}q^{-N+1};q)_n}
\sum_{k=0}^n\frac{(q^{-n};q)_k(a^2q^m;q)_k(q^{-m};q)_k q^k}
{(ab;q)_k(q^{-N+1};q)_k(q;q)_k}=\\
(z;q)_m\sum^{N-1-m}_{k=0}\frac{(q^{m-N+1};q)_k(abq^m;q)_k q^{-mk} z^k}
{(ba^{-1}q^{-N+1};q)_k (q;q)_k a^{2k}}=\\
(z;q)_m P_{N-1-m}(za^{-2}q^{-m-1},ba^{-1}q^{-N},a^2q^{2m}|q),}
\label{eigenvecq2}
\end{equation}
where $P_k(x,\alpha,\beta|q)$ are the little $q$-Jacobi
polynomials. Thus the equation (\ref{deq2}) is related to the $q$-difference
equation for these polynomials [\ref{Koekoek}] by the transformation
$ P_{N-1-m}(x)=f_m(xa^2q^{m+1})/(xa^2q^{m+1};q)_m $.

Finally, consider the dual q-Hahn polynomials. (Other known basic 
hypergeometric polynomials leading by
the procedure of this section to equations of the type (2) can be considered
as particular cases of the continuous dual q-Hahn or dual q-Hahn
polynomials.) These polynomials are defined by the recurrence relation 
(we use $N-1$ instead of $N$ in the usual definition [\ref{Koekoek}])
\begin{equation}
\eqalign{
\mu(y) p_n=A_np_{n+1}+(1+\gamma\delta q -A_n-C_n)p_n+C_np_{n-1},\\
A_n=(1-q^{n-N+1})(1-\gamma q^{n+1}),
\qquad C_n=\gamma q(1-q^n)(\delta-q^{n-N}),\\
\mu(y)=q^{-y}+\gamma\delta q^{y+1},\qquad p_{-1}=0,
\qquad n=0,1,\dots,N-1.}\label{dqh}
\end{equation}
Setting $\gamma=abq^{-1}$, $\delta=ab^{-1}$, $q^{-N+1}=ac$,
and multiplying the recurrence relation (\ref{dqh}) by $a^{-1}$, we obtain 
(\ref{r1q}) where $2x=t+t^{-1}$, $t=aq^{y}$.
Thus, the first $N$ continuous dual q-Hahn polynomials at $ac=q^{-N+1}$ and 
the dual q-Hahn polynomials are the same (up to renaming the parameters). 

\section{Acknowledgements}
I am grateful to A. V. Turbiner for many valuable suggestions and discussions.
I also thank Ya.~I.~Granovskii, L.~L.~Vaksman, and A.~V.~Zabrodin for useful 
discussions.

\end{document}